\def\year{2016}
\documentclass[letterpaper]{article}
\usepackage{aaai16}
\usepackage{helvet}
\usepackage{courier}
\usepackage{helvet}
\usepackage{courier}
\usepackage{balance}  
\usepackage{graphics} 
\usepackage{times}    
\usepackage[usenames,dvipsnames]{xcolor}
\usepackage{booktabs}
\usepackage{graphicx}
\usepackage{tabularx}
\usepackage{dcolumn}
\usepackage{amsmath}

\begin{document}

\title{Two Tales of the World: Comparison of Widely Used World News Datasets GDELT and EventRegistry \\{\LARGE [Please cite the ICWSM'16 version of this paper]}}

\author{Haewoon Kwak \quad Jisun An\\
Qatar Computing Research Institute\\
Hamad Bin Khalifa University\\
\{hkwak, jan\}@qf.org.qa}
\maketitle

\begin{abstract}
In this work, we compare GDELT and Event Registry, which monitor news articles worldwide and provide big data to researchers regarding scale, news sources, and news geography.  We found significant differences in scale and news sources, but surprisingly, we observed high similarity in news geography between the two datasets.
\end{abstract}

\section{Introduction}

The philosophy of computational journalism has been shaping a new research direction on journalism.  Research that usually involves survey participants is now being conducted by using large-scale data.  

One of the outstanding efforts in this research area is the GDELT project\footnote{http://www.gdeltproject.org/}, which monitors ``the world's broadcast, print, and web news from nearly every corner of every country in over 100 languages''~\cite{leetaru2013gdelt}. GDELT translates all other languages into English through a collaboration with Google Ideas, and adds metadata, such as a location where an event happens, to each news article. A vast array of studies from prediction~\cite{heaven2013world} to analysis~\cite{kwak2014first} and comparison with private data~\cite{ward2013comparing} have been published to explore the potential of the dataset.

On the other hand, recently, a new platform called Event Registry (ER)\footnote{http://eventregistry.org/} has been launched. It offers a large collection of news articles through the public API~\cite{rupnik2015news}. 

In this work, we compare GDELT and ER in terms of scale, news sources, and news geography. Given that there is no ground truth of world news datasets, we believe that the best effort to understand the benefit and limitation of the existing datasets is a comparison of the two publicly available datasets.  

Two datasets showed stark differences in terms of the number of news articles and sources being indexed. Thus, we suggest that the dataset be used with caution.  However, at the same time, we found that news geographies drawn from two datasets were extremely similar, indicating that the overall trend of international news coverage seems to be consistent.

\section{Data Collections}

Given that there is a rate limit in using Event Registry, it is infeasible to compare the entire datasets of two services. Thus, we sample the two datasets.  We use the first day of every month from March to December 2015, to avoid temporary fluctuations.  We exclude the first day of January and February because GDELT v2.1 has been available since February 18, 2015. 

We collect all compressed dump files for English\footnote{http://data.gdeltproject.org/gdeltv2/masterfilelist.txt (Last access: 5 Jan 2016)} and Translingual\footnote{http://data.gdeltproject.org/gdeltv2/masterfilelist-translation.txt (Last access: 5 Jan 2016)} during the 10 target days. For each day, GDELT releases 4 $\times$ 24 $\times$ 2 = 192 files (one file each every 15 minutes for English and Translingual).  As a result, we collect 192 $\times$ 10 = 1920 compressed files from the GDELT project.  

Unlike GDELT, Event Registry offers API to access their data.  We query the indexed articles in  the first day of each month by using \textit{QueryArticles().setDateLimit()} and collect all of them with pagination. We also query all the details of the articles and sources.  We upload the code snippet in Github Gist\footnote{https://gist.github.com/anonymous/943574147ffaf3b1b2f4}.

\section{Comparison between GDELT and ER}

\subsection{Scale}
 
\begin{figure} [hbt!]
  \begin{center}
  \includegraphics[width=70mm]{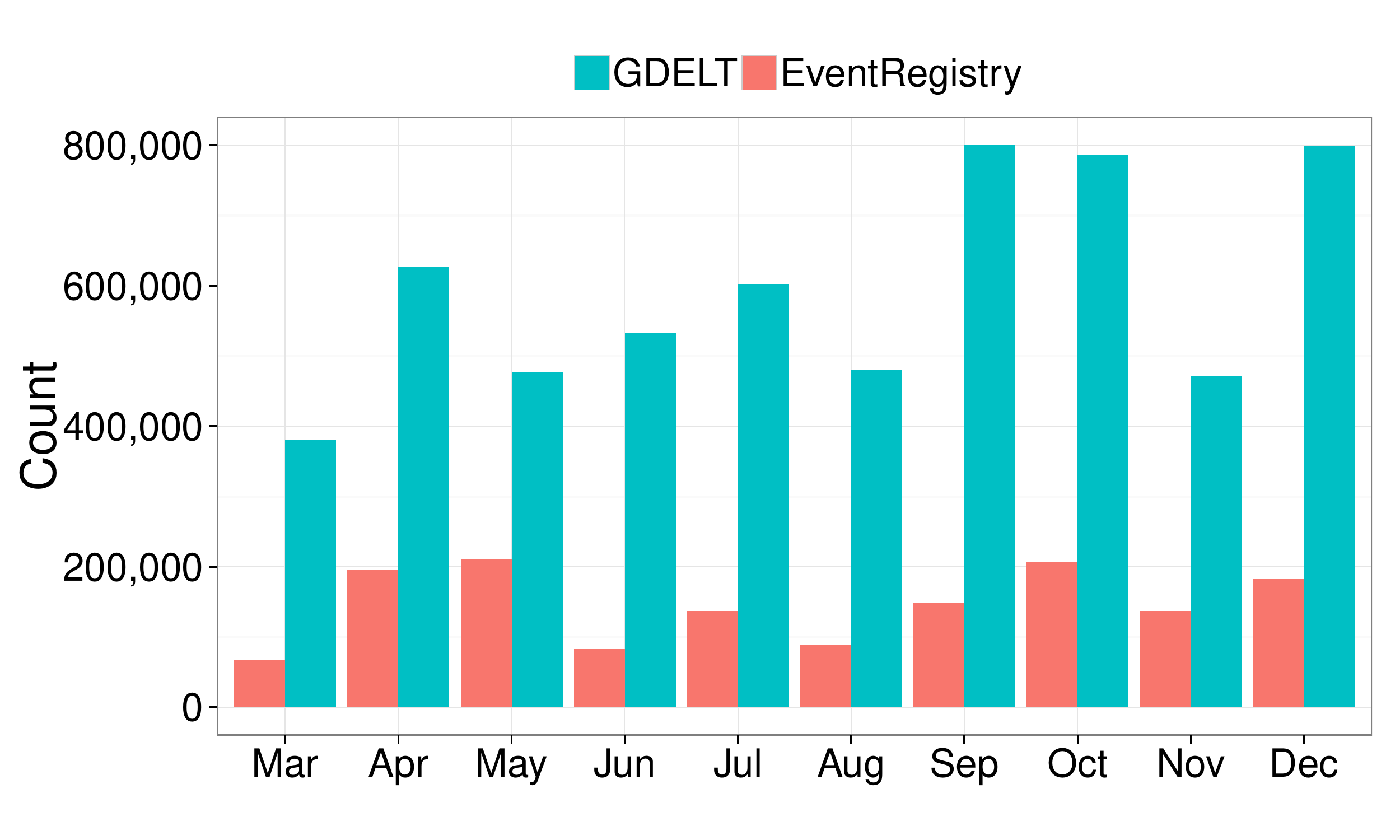}
  \caption{Number of indexed articles in both platforms}  
  \vspace{-1mm}
  \label{fig:monthly_articles}
  \end{center}
\end{figure}

One of the most important criteria for the comparison is the scale of a dataset because it describes how comprehensive the dataset is. Figure~\ref{fig:monthly_articles} shows the number of articles indexed by the two platforms on the first day of each month from March to December 2015. The daily volumes of news articles over time are highly fluctuating in both datasets. Such fluctuation is mainly because news media publishes a different number of articles on weekdays and weekends. All three lower bars in the GDELT dataset represent the number of indexed articles on weekends $[$March (Sunday), August (Saturday), and November (Sunday)$]$. While the plot representing ER does not entirely follow the ``high in weekday, low in weekend'' pattern, a positive correlation exists between two time-series from GDELT and ER (Pearson correlation coefficient $r$=0.57, p=0.08). 

Another interesting observation is that GDELT consistently indexes 2.26 times (May) to 6.43 times (June) more articles than ER, even with high fluctuations over time. To better understand where this significant difference comes from, we quantitatively examine the sources of the articles indexed by GDELT. GDELT explicitly announces that it collects documents not only from the world wide web but also from ``broadcast, print, or other offline sources,'' whereas ER mainly focuses on news articles through RSS feeds. Thus, having more sources might be the reason why GDELT indexes much more documents than ER. 

However, we find that 99.9\% of the documents are collected from the web, and only less than 1,000 documents in each day are from other offline sources.  Moreover, all the documents that are not from the web are collected from BBC Monitoring, which monitors mass media worldwide, and no other individual source is found.  That is, the primary data source for both platforms is the web, although GDELT claims that they collect documents from various sources. Also, the significant difference in the number of indexed articles between the two platforms is solely based on the difference in the news coverage of different sources.

\begin{table}[t!]
\begin{center}
\footnotesize \frenchspacing
\begin{tabular}{ccrcr}
\toprule
Rank & \multicolumn{2}{c}{GDELT} & \multicolumn{2}{c}{{ER}}  \\
\midrule
1 & English & 2,415,060 & English & 765,637 \\
2 & Spanish & 646,052 & Spanish & 164,487 \\
3 & Arabic & 320,085 & German & 130,882 \\
4 & German & 301,297 & French & 83,180 \\
5 & Turkish & 259,364 & Russian & 79,678 \\
\midrule
6 & French & 242,082 & Portuguese & 55,747 \\
7 & Chinese & 232,795 & Chinese & 44,717 \\
8 & Russian & 230,005 & Italian & 37,214 \\
9 & Italian & 169,324 & Arabic & 34,928 \\
10 & Portuguese & 143,288 & Turkish & 30,941 \\
\bottomrule
\end{tabular}
\end{center}
\vspace{-2mm}
\caption{Top 10 languages of documents}
\vspace{-3mm}
\label{tbl:top_languages}
\end{table}

\subsection{Language}
We look into account the language of the indexed articles to compare the coverage of the two datasets. We find that the GDELT indexes documents written in 64.1 different languages on average (across 10 different sample datasets), whereas ER indexes articles of only 14 languages. For both platforms, the most dominant language is English. Given that GDELT covers more languages, the proportion of English articles in GDELT (2,415,060 articles, 40.6\%) is naturally lower than in ER (765,637 articles, 51.8\%).

Then, what other languages do they index?  Table~\ref{tbl:top_languages} shows the top 10 languages in both platforms.  While there are some differences in rankings, the 10 languages are the same. Interestingly, they are not the same as the top 10 languages by the number of native speakers\footnote{https://goo.gl/jjUxj1}, total speakers\footnote{https://goo.gl/VNBTcZ}, nor Internet web pages\footnote{https://goo.gl/c7Xm6l}.

\subsection{News Sources}
Next, we count the number of unique news sources (website domains) for each language and compute the correlation between this source and the number of documents written in each language. We find a significant positive Spearman correlation coefficient for GDELT ($\rho$=.91, p $<$ .0001) and  ER ($\rho$=.94, p $<$ .0001). This high correlation is trivial because more news sites naturally lead to more news articles to index. 

\begin{table}[t!]
\begin{center}
\small \frenchspacing
\begin{tabular}{ccr}
\toprule
Language & Domain & Articles \\
\midrule
spa & www.entornointeligente.com & 35,932 \\
deu & www.gaeubote.de & 27,212 \\
fra & www.lopinion.ma & 23,610 \\
deu & www.krankenkassen-direkt.de & 22,399 \\
fra & www.fasopresse.net & 17,314 \\
\midrule
spa & www.surenio.com.ar & 15,243 \\
eng & www.4-traders.com &  14,554 \\
zho & news.sina.com.tw  &  14,509 \\
ita & www.agenzianova.com & 13,261 \\
tur & www.haberler.com  &  12,575 \\
\bottomrule
\end{tabular}
\end{center}
\caption{Top 10 domains in ER}
\label{tbl:top_domain_er}
\end{table} 

\begin{table}[t!]
\begin{center}
\small \frenchspacing
\begin{tabular}{ccr}
\toprule
Language & Domain & Articles \\
\midrule
eng & www.dailymail.co.uk & 10,051 \\
eng & www.reuters.com & 9,137 \\
spa & www.eleconomista.es & 6,058 \\
eng & www.bizjournals.com & 5,875 \\
tur & www.merhabahaber.com    & 5,738 \\
\midrule
eng & www.prnewswire.com  & 5,516 \\
zho & udn.com & 5,279 \\
eng & www.nydailynews.com & 4,322 \\
zho & www.chinanews.com   & 4,306 \\
eng & hosted.ap.org   & 4,271 \\
\bottomrule
\end{tabular}
\end{center}
\caption{Top 10 domains in GDELT}
\vspace{-3mm}
\label{tbl:top_domain_gdelt}
\end{table} 

\begin{figure} [hbt!]
  \begin{center}
  \vspace{-2mm}  
  \includegraphics[width=70mm]{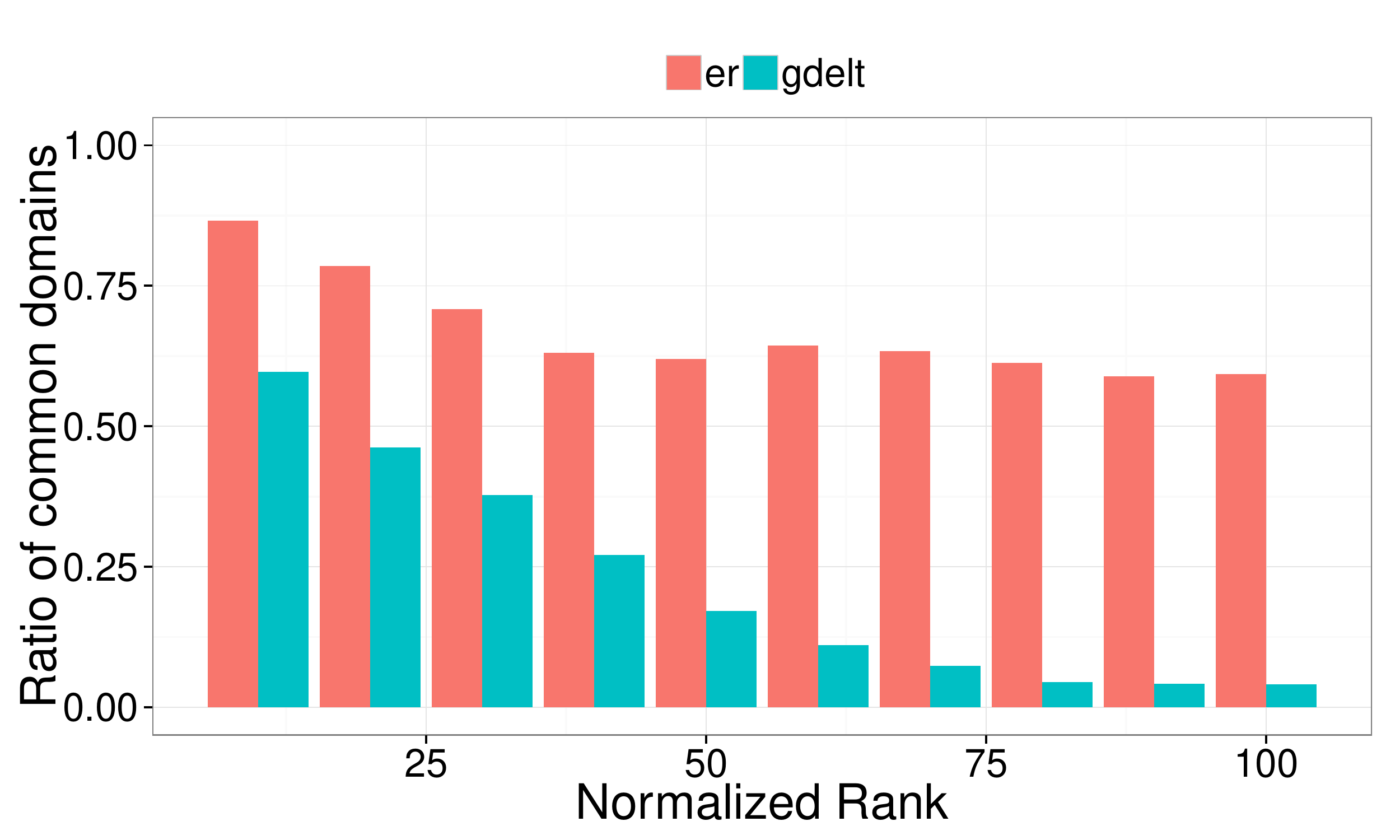}
  \caption{Proportion of common domains according to its rank}  
  \label{fig:overlap}
  \vspace{-1mm}
  \end{center}
\end{figure}

We look into the unique domains in each dataset. Among 20,754 domains in ER and 63,268 domains in GDELT, 13,867 domains are common. We examine whether any pattern exists among the common domains.  We first rank the domains by the number of articles that are published in each of them.  We then compute how the proportion of common domains between two platforms change from the top 10\% of domains (leftmost bar in Figure~\ref{fig:overlap}) to the bottom 10\% of domains (rightmost bar in Figure~\ref{fig:overlap}).  

Figure~\ref{fig:overlap} shows the proportion of common domains between the two platforms according to normalized rank. In GDELT, we see a clear pattern that a more active domain is likely to be common, and a less active domain tends to be found only in GDELT. However, in ER, this pattern is weak.  The pattern holds only up to the top 40\%, but the proportion of common domains does not vary much after that point. Moreover, it is surprising that ER does not index 40.3\% of the top 10\% of domains in GDELT.  More than a half of top 10\% to 20\% of domains in  GDELT are not indexed by ER.  On the other hand, GDELT covers more of ER's domains.  Notably, 13.4\% of the top 10\% of domains in ER and 21.5\% of the top 10\% to 20\% of domains in ER are not indexed by GDELT. This finding shows that, even though the number of unique websites indexed by ER is one-third of those indexed by GDELT, ER is not a simple subset of the top websites of GDELT. Rather, both platforms have unique sets of news sources independently, while having some common news sources.  So it is likely that news coverage research based on each platform might show somewhat different results. Thus, studies are required to characterize the differences between news sites indexed by both platforms and find a way to address them.

We further examine the top sources (domains) of two datasets. Table \ref{tbl:top_domain_er} reports the top 10 ranked list of news sources in GDELT and \ref{tbl:top_domain_gdelt} reports that in ER. Firstly, we see that there is no overlap between the two lists. Then, we find the  few, top news sources have an extremely large number of articles in the two datasets compared to other sources. The average number of articles per source is 89.97 in GDELT and 67.60 in ER, and the median is 4 in GDELT and 6 in ER. Compared to these values, the top 10 biggest sources in Table \ref{tbl:top_domain_er} and \ref{tbl:top_domain_gdelt} are extreme.

\begin{figure} [hbt!]
  \begin{center}
  \includegraphics[width=\columnwidth]{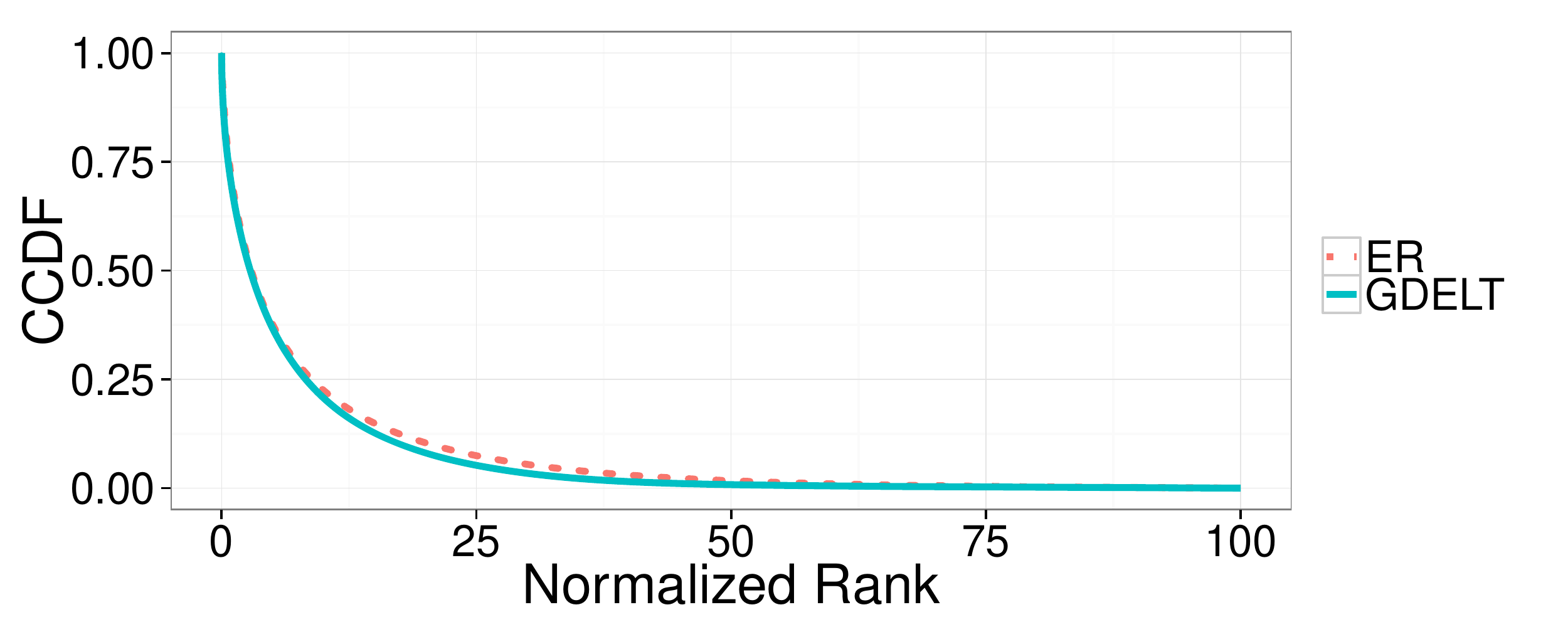}
  \vspace{-3mm}
  \caption{Proportion of articles from top x\% of sources}  
  \label{fig:ranked_CCDF}
  \vspace{-1mm}
  \end{center}
\end{figure}

Figure~\ref{fig:ranked_CCDF} shows the dominant proportion of the top sources.  News articles from the top 10\% of sources account for 77.5\% of the whole set of collected articles in ER and 79.2\% of those in GDELT.  This finding warns researchers to design experiments to see news coverage and interpret the result when using either of the two datasets.  In contrast to offline news media that have limited resources (e.g., one-hour broadcasting news or 12'' $\times$ 23'' paper size of a newspaper) for publishing news, online news media have no physical limitations,  and it costs less than print to publish news articles.  This means that the number of news articles published by online news media might have less significant meaning compared to offline news media. Therefore, it requires a careful normalization when taking into account the number of articles published by each news source. Otherwise, the core measure in news coverage studies, ``how many articles report a target incident,'' could be biased by the behaviors of the dominant news media sources.

Again, an appropriate normalization technique should be carefully selected. In our case, the market share of each news medium, which indicates its reach to an audience, is typically considered as one of the prominent factors to infer its importance.  None of these platforms offer such metrics.  While ER is supposed to provide a field of ``importance'' in the source, its value is currently set to zero equally for all the news sources.

As a preliminary guideline, we test the method to infer the importance of each website based on its traffic, as used in the study by \cite{kwak2014first}.  We collect traffic information from the top 100 media sites in GDELT and ER from Alexa.com.  Alexa shows the global rank of each website based on its traffic.   
We compute the Spearman correlation coefficients between the global rank of Alexa and the rank in both platforms.  We find almost no correlation in GDELT ($\rho$=-.009) and an insignificant, weak positive correlation in ER ($\rho$=.17, p=.10).  This finding suggests that the number of articles published by a source should not be used as a proxy for its important or reach to an audience.

\subsection{News Geography}

News geography is the extent to which countries are reported in the news internationally~\cite{sreberny1984world}.  Typically, the number of news articles mentioning a country is used as a measure of the news geography. The measure shows the selection bias by journalists, and it is widely used for research on foreign news coverage~\cite{wilke2012geography,kwak2014first}.  

GDELT extracts all the locations from the text and stores them in the ``V2ENHANCEDLOCATIONS'' field. Similarly, ER extracts all the concepts from the text and adds metadata to each concept.  We extract concepts that have a valid value in the ``location'' and ``country'' fields.

\begin{figure} [hbt!]
  \begin{center}
\vspace{-9mm}  
  \includegraphics[width=\columnwidth]{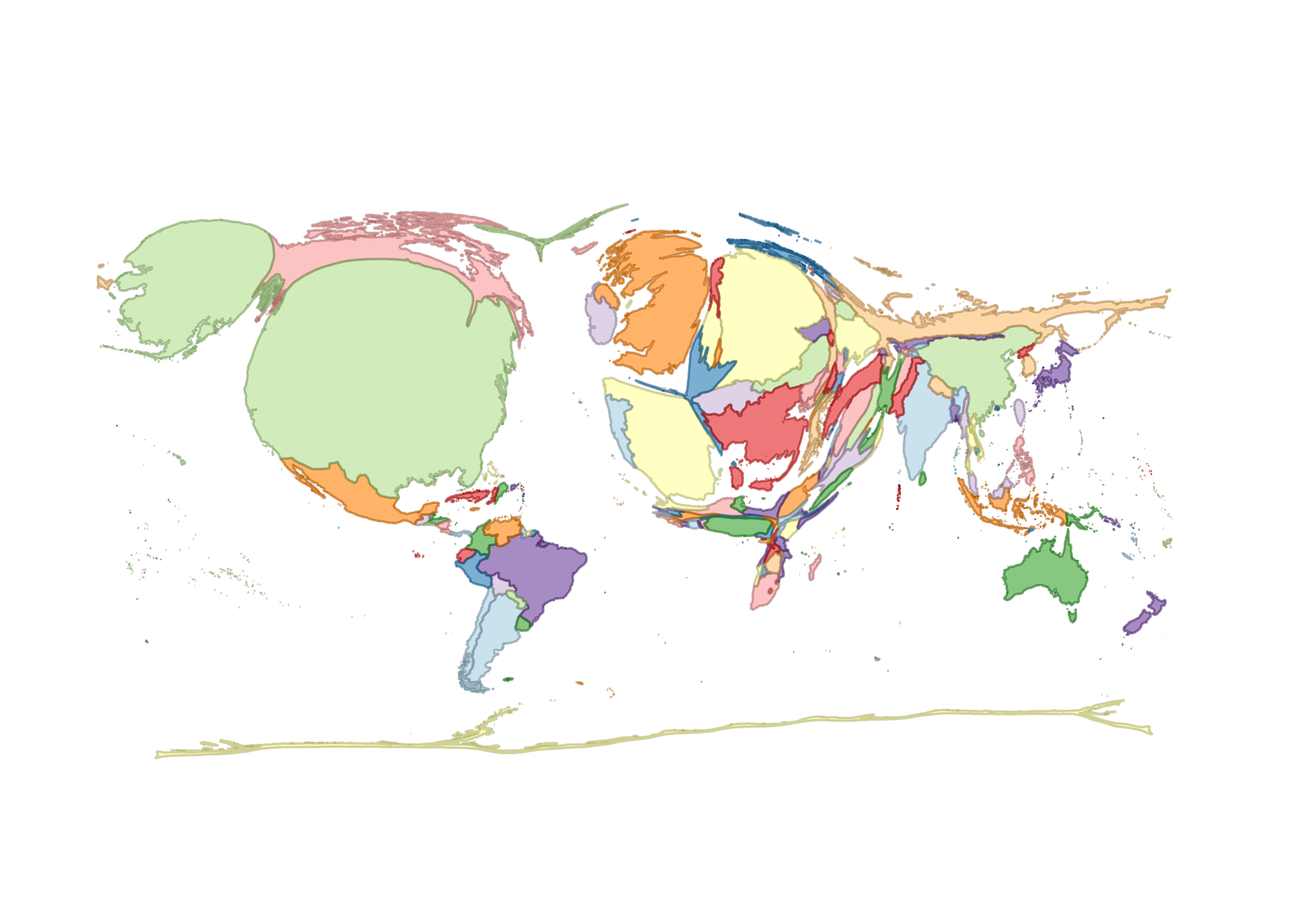}
  \vspace{-13mm}
  \caption{News geography of ER dataset}  
  \label{fig:er_ng}
  \end{center}
  \vspace{-1mm}
\end{figure}

Figure~\ref{fig:er_ng} shows the news geography of ER.  The size of the territory of a country is proportional to the number of articles mentioning the country. We observe that the United States, United Kingdom, Germany, and Spain are overrepresented compared to their original territory.  That is, ER contains more news articles mentioning events happening in those countries. The news geography of GDELT is similar, as we see in Figure~\ref{fig:er_ng} but China and Russia are represented more. We have omitted them due to lack of space.

$D^{E}_{C_i}$ denotes the documents about country $C_i$ in ER, and $D^G_{C_j}$ denotes those about country $C_j$ in GDELT.  Then, we can define the news geography of ER and GDELT as an ordered list of the number of documents about country $C_i$ ($i$ = 1..$n$) in ER and GDELT, respectively: 
\begin{align*}
NG^E=[|D^{E}_{C_1}|, |D^{E}_{C_2}|, ..., |D^{E}_{C_n}|]\\
NG^G=[|D^{G}_{C_1}|, |D^{G}_{C_2}|, ..., |D^{G}_{C_n}|]
\end{align*}

We can then measure the similarity between the news geography of both datasets by using corr($NG^E$, $NG^G$).  
We find a significantly high correlation between the news geographies of ER and GDELT ($\rho$=0.867, $p$=1.896e-74). 

This finding is particularly surprising. In the previous section, we observe that the news sources indexed by both datasets are largely different, but what they regard as \textit{news} are similar in terms of where an event happens.  Our finding is adding up another evidence that journalists have a concept of ``world hierarchy''~\cite{chang1998all} in determining which is more important foreign news.

\section{Summary and Discussion}

In this work, we compared two publicly available datasets on world news in terms of scale, news sources, and news geography. We found significant differences in scale and news sources, but they are similar in news geography.

We found that GDELT collects 2.26 times to 6.43 times more documents than ER does per day. GDELT indexes documents in 64.1 different languages per day on average, whereas ER indexes documents in 14 languages.  The top 10 languages in both datasets are the same, with only subtle differences in ranking.  However, these top 10 languages are not the same as those by the number of total speakers nor those by the language commonly used on the Internet. 

We also discovered that GDELT indexes documents from 63,268 websites, and ER from 20,754 websites.  Among them, 13,867 websites are common.  We observe a trend that the websites with the biggest number of published documents are likely to be common, but at the same time, each dataset has a lot of unique websites publishing many documents.  

The number of articles from a news source varies greatly across the sources.  The top 10\% of news sources publish about 80\% of news articles, collected in both datasets.  Also, we find that the number of documents from a source does not correlate with the web traffic of the corresponding source.  

Finally, in spite of all the differences between the two datasets, we found that news geography obtained from both datasets are quite similar.  

What we learned from this study is that we should carefully use GDELT and ER for research because the two datasets are quite different in terms of scale and news sources.  It is essential for researchers to understand the data accurately prior to a deeper analysis of either dataset.  This also encourages the sharing of the dataset (or, a set of document identifiers, at least) used in research. 

Another practical consideration for research using both datasets is accessibility.  In contrast to GDELT, which allows researchers to download the compressed raw files, ER provides the API with a rate limit.  Even though we tried to maximize the size of the response, it took a few days to collect the articles used in this study.  

For future research, we would collect additional information about news sources and propose a method to use both datasets together, considering the importance of each news source. 

\footnotesize
\bibliographystyle{aaai}
\bibliography{icwsm2016-gdelt-main}
\end{document}